\documentclass[a4paper,fleqn,usenatbib]{mnras}

\usepackage[T1]{fontenc}
\usepackage{ae,aecompl}

\usepackage{graphicx}	
\usepackage{amsmath}	
\usepackage{amssymb}	
\usepackage{subfigure}
\usepackage{multirow}
\usepackage{longtable}

\newcommand{\pcc}{\,{\rm cm}^{-3}}
\newcommand{\gcc}{\,{\rm g \, cm}^{-3}}

\newcommand{\um}{\, {\rm \mu m}}
\newcommand{\nm}{\, {\rm nm}}

\newcommand{\kel}{\, {\rm K}}

\newcommand{\nh}{n_{\rm H}}

\newcommand{\amin}{a_{\rm min}}
\newcommand{\amax}{a_{\rm max}}

\newcommand{\mh}{m_{\rm H}}

\newcommand{\yr}{\, {\rm yr}}
\newcommand{\kms}{\, {\rm km \, s^{-1}}}
\newcommand{\ssi}{_{\rm Si}}
\newcommand{\depsi}{{\rm [Si/H]_{gas}}}

\newcommand{\myr}{\, {\rm Myr}}

\title[Diffuse grain growth]{The efficiency of grain growth in the diffuse interstellar medium}

\author[Priestley et al.]{
  F. D. Priestley$^{1}$\thanks{Email: priestleyf@cardiff.ac.uk},
  I. De Looze$^{2,3}$ and
M. J. Barlow$^{3}$
\\
$^{1}$School of Physics and Astronomy, Cardiff University, Queen's Buildings, The Parade, Cardiff CF24 3AA, UK \\
$^{2}$Sterrenkundig Observatorium, Ghent University, Krijgslaan 281 - S9, 9000 Gent, Belgium\\
$^{3}$Department of Physics and Astronomy, University College London, Gower Street, London WC1E 6BT, UK\\
}

\date{Accepted XXX. Received YYY; in original form ZZZ}

\pubyear{2021}

\begin{document}
\label{firstpage}
\pagerange{\pageref{firstpage}--\pageref{lastpage}}
\maketitle

\begin{abstract}

  Grain growth by accretion of gas-phase metals is a common assumption in models of dust evolution, but in dense gas, where the timescale is short enough for accretion to be effective, material is accreted in the form of ice mantles rather than adding to the refractory grain mass. It has been suggested that negatively-charged small grains in the diffuse interstellar medium (ISM) can accrete efficiently due to the Coulomb attraction of positively-charged ions, avoiding this issue. We show that this inevitably results in the growth of the small-grain radii until they become positively charged, at which point further growth is effectively halted. The resulting gas-phase depletions under diffuse ISM conditions are significantly overestimated when a constant grain size distribution is assumed. While observed depletions can be reproduced by changing the initial size distribution or assuming highly efficient grain shattering, both options result in unrealistic levels of far-ultraviolet extinction. We suggest that the observed elemental depletions {in the diffuse ISM} are better explained by {higher initial depletions, combined with inefficient} dust destruction by supernovae at moderate ($\nh \sim 30 \pcc$) densities, rather than by higher accretion efficiences.

\end{abstract}




\section{Introduction}

The growth of dust grains by accretion of gas-phase metals is widely assumed to be an important process in the interstellar medium (ISM). Observations such as the correlation of depletion with gas density \citep{jenkins2009} are naturally explained by this mechanism, and studies of galaxy evolution have typically found it a necessary addition to models in order to reproduce both present-day \citep{popping2017,devis2019,triani2020} and high-redshift \citep{mancini2015,graziani2020} dust masses, although some recent works have challenged this \citep{gall2018,delooze2020,nanni2020}. Despite this, our understanding of how grain growth actually works is limited. In dense gas, where accretion is efficient, dust grains form icy mantles rather than directly accreting refractory elements, which are rapidly photodesorbed when exposed to ultraviolet (UV) radiation \citep{barlow1978}. \citet{ferrara2016} and \citet{ceccarelli2018} have argued that this makes any actual increase in bulk dust mass via accretion impossible.

\citet{zhukovska2016,zhukovska2018} identified a potential way to avoid this issue by considering grain growth in the cold neutral medium (CNM). While the low density makes accretion highly inefficient for traditional models (e.g. \citealt{hirashita2011}), under these conditions sufficiently small dust grains become negatively charged \citep{weingartner2001}. As most dust-forming elements exist as singly-charged positive ions when exposed to UV radiation, Coulomb attraction can significantly enhance the accretion rate \citep{weingartner1999}, leading to much more efficient grain growth than otherwise expected. \citet{zhukovska2016,zhukovska2018} implemented this mechanism into a hydrodynamical simulation of the ISM, finding that the observed patterns of silicon and iron depletion with respect to gas density were well-reproduced, although requiring a somewhat smaller minimum grain size than the typical \citet{mathis1977} (MRN) distribution.

The models in \citet{zhukovska2016,zhukovska2018} assumed a constant power law grain size distribution, the properties of which are treated as input parameters. However, grain growth, by definition, involves an evolution of the size distribution. This results in small grains rapidly becoming significantly larger \citep{hirashita2011}, to the point where the \citet{weingartner1999} model predicts they become positively charged and repel, rather than attract, positive ions. In this paper, {we show that the inclusion of a consistently evolving size distribution significantly reduces the effiency of grain growth in the diffuse ISM. Observed elemental depletions in the CNM can only be reproduced by assuming either very efficient grain shattering, or a further reduction in the minimum grain size from the \citet{zhukovska2016} model, both of which result in extinction laws in conflict with Galactic measurements. We suggest that the assumption of rapid dust destruction by supernovae in the CNM is at fault. Relaxing this assumption, the observed depletion patterns can be reproduced as long as the material injected into the CNM is already significantly depleted ($\depsi \sim -1.5$).}

\section{Method}

\begin{figure}
  \centering
  \includegraphics[width=\columnwidth]{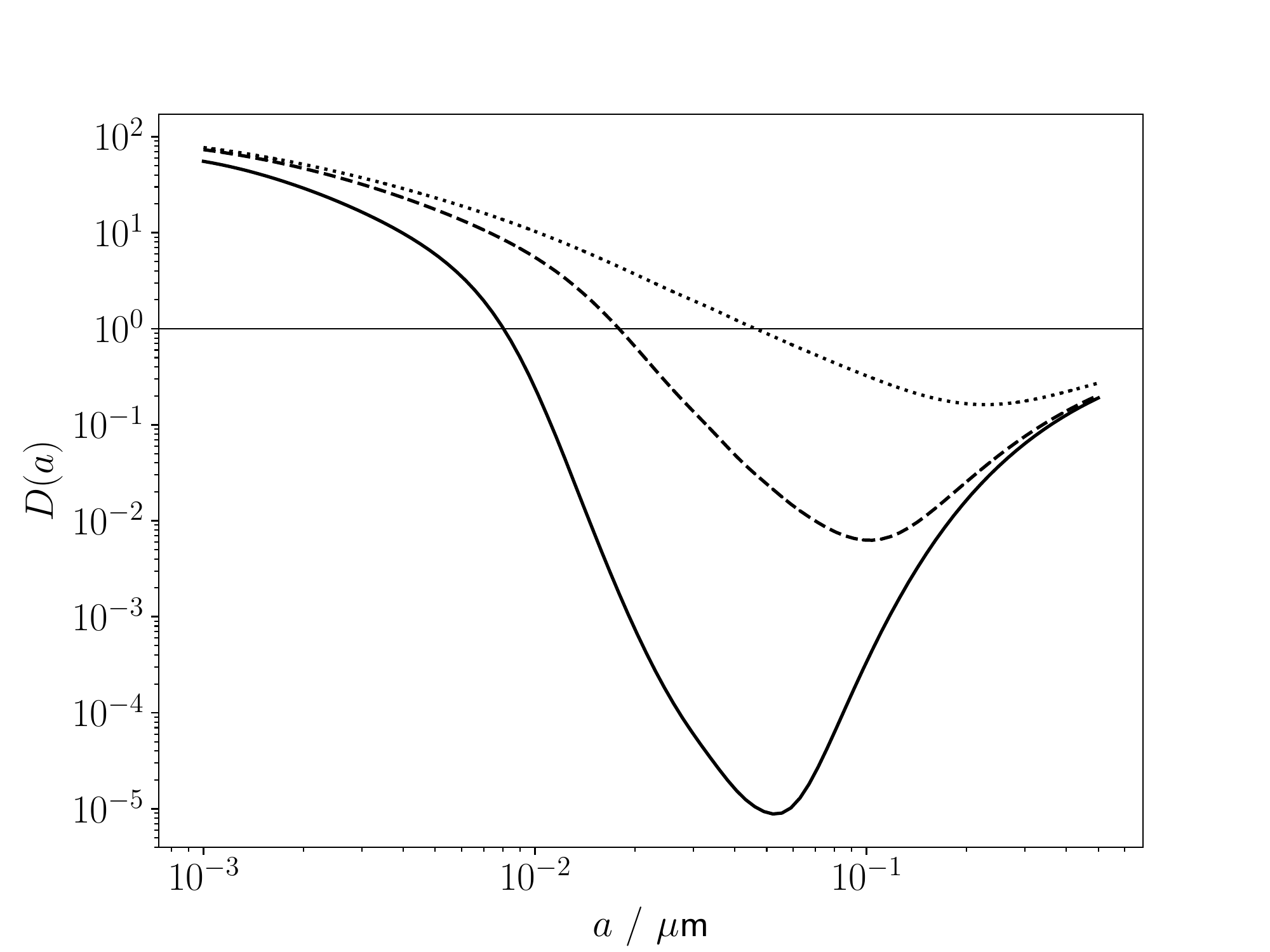}
  \caption{Coulomb focusing factor versus grain size for $T=100 \kel$, $x_e = 0.0015$ and $\nh = 5$ (solid line), $30$ (dashed line) and $100 \pcc$ (dotted line). The thin solid line marks $D(a) = 1$, i.e. no net attraction or repulsion.}
  \label{fig:coulomb}
\end{figure}

\begin{table}
  \centering
  \caption{Model parameters}
  \begin{tabular}{ccc}
    \hline
    Parameter & Value & Unit \\
    \hline
    Gas density $\nh$ & $30$ & $\pcc$ \\
    Gas temperature $T$ & $100$ & $\kel$ \\
    Electron fraction $x_e$ & $0.0015$ & - \\
    Grain density $\rho$ & $3.13$ & $\gcc$ \\
    Silicon mass fraction $f\ssi$ & $0.165$ & - \\
    Silicon elemental abundance $\epsilon\ssi$ & $3.24 \times 10^{-5}$ & - \\
    Initial depletion $\depsi$ & $-0.5$ & - \\
    Dust destruction timescale $\tau_{\rm dest}$ & $350$ & $\myr$ \\
    Maximum grain size $\amax$ & $0.25$ & $\um$ \\
    Minimum grain size $\amin$ & $0.005$ & $\um$ \\
    \hline
  \end{tabular}
  \label{tab:params}
\end{table}

We follow, as closely as possible, the formalism and parameter values used by \citet{zhukovska2016}. We consider silicate grains, with a density of $\rho = 3.13 \gcc$ and a silicon mass fraction $f\ssi = 0.165$, and assume that growth is limited by the availability of silicon atoms. We take the elemental abundance of silicon to be $\epsilon\ssi = 3.24 \times 10^{-5}$ \citep{lodders2009}. In the absence of grain charge, the rate of change of grain radius is then
\begin{equation}
  \frac{da}{dt} = \frac{n\ssi m\ssi <v\ssi>}{4 \rho f\ssi}
  \label{eq:gr}
\end{equation}
where $n\ssi$, $m\ssi = 28 \mh$ and $<v\ssi>$ are the gas-phase number density, mass and mean thermal velocity of silicon atoms respectively. In principle, this equation may be modified by a sticking probability, dependent on gas/grain temperature among other properties. \citet{zhukovska2016} find different implementations of the sticking probability have little impact on their results, so we assume the sticking probability is unity under CNM conditions.

Equation \ref{eq:gr} is independent of grain radius. The introduction of grain charge, which varies with size, modifies the growth rate by a Coulomb focusing factor $D(a)$ \citep{weingartner1999}, the ratio of the actual cross-section for collision to the geometric cross section. We calculate the grain charge distributions for sizes between $0.001-0.5 \um$ following \citet{weingartner2001}, and convert these into focusing factors assuming singly-charged positive ions using the formulae presented in \citet{weingartner1999}. Figure \ref{fig:coulomb} shows the variation of $D(a)$ with grain size for a range of densities, for gas with temperature $T = 100 \kel$ and electron fraction $x_e = 0.0015$. For typical CNM conditions, small grains tend to be negatively charged and thus attract positive ions, but beyond a radius of $\sim 0.01 \um$ grains instead become positively charged and $D(a) < 1$. This transition radius increases with density as electron attachment becomes more effective (for constant $x_e$), which also reduces the magnitude of the repulsion effect, but the qualitative behaviour is the same.

We note that while \citet{weingartner1999}, and by extension \citet{zhukovska2016,zhukovska2018}, assume an electron fraction in the CNM of $0.0015$, this requires the majority of electrons to come from hydrogen or helium, and including grain-assisted recombination for these elements has been shown to reduce the CNM electron fraction to $\sim 10^{-4}$ \citep{weingartner2001b,liszt2003}. We explore the effect of this lower value in Appendix \ref{sec:lowe}, finding that it significantly reduces the accretion efficiency due to generally more positively-charged grains. Nonetheless, we continue to use the higher value for consistency with previous work.

We assume an initial MRN grain size distribution with $\amin = 0.005 \um$, $\amax = 0.25 \um$ and a power law index of $-3.5$, and an initial silicon depletion value $\depsi = \log_{10} \left( \frac{n\ssi}{\epsilon\ssi \nh} \right) = -0.5$, again following \citet{zhukovska2016}. The initial $\depsi$ allows us to determine the initial number density of grains, $n_g(a)$. We divide the size distribution into 100 logarithmically-distributed bins, and use Equation \ref{eq:gr} combined with $D(a)$ to calculate the growth rate for each size bin. For a time interval $dt$ we can then calculate the increase in grain radius, $da$, and the new grain size, $a_1 = a_0 + da$. As we later implement processes which do not conserve grain number, rather than simply updating $a$ for each bin, we redistribute the grains in bin $i$ between bins $j$ and $j+1$ where the new radius $a_1$ falls between $a_j$ and $a_{j+1}$, such that
\begin{equation}
  dn_g(j) = n_g(i)\left(1 - \frac{a_1 - a(j)}{a(j+1)-a(j)}\right)
\end{equation}
and
\begin{equation}
  dn_g(j+1) = n_g(i)\left(1 - \frac{a(j+1)-a_1}{a(j+1)-a(j)}\right).
\end{equation}
The new mass in grains due to the increase in total volume is then used to update the remaining gas-phase abundance of silicon $n\ssi$.

Note that as we used a fixed grid of grain sizes, the largest grain size does not grow, and over time smaller grains `pile up' in the last bin as they reach this size. For an MRN distribution, the accretion rate is always dominated by the smallest grain sizes, particularly when charge is included (as large grains are positively charged and so have $D(a) < 1$). Even without grain charge, the difference in the mass accreted when accounting for this effect is negligible, as the fractional increase in grain radius (and thus volume) for the largest grains is tiny. In any case, this has little effect on our overall argument.

\citet{zhukovska2016} model the growth of dust mass via the equation
\begin{equation}
  \frac{df_d}{dt} = \frac{1}{\tau_{\rm acc}} f_d (1 - f_d)
  \label{eq:gr2}
\end{equation}
where $f_d$ is the fraction of silicon locked up in dust grains (i.e. $f_d = 1 - 10^\depsi$) and $\tau_{\rm acc}$ is the accretion timescale, defined as
\begin{equation}
  \tau_{\rm acc}^{-1} = \frac{3 \epsilon\ssi m\ssi v\ssi \nh}{\rho f\ssi <a>_3}.
  \label{eq:tscale}
\end{equation}
The quantity $<a>_3$ is the average grain radius accounting for Coulomb focusing, given by
\begin{equation}
  <a>_3 = <a^3>/<D(a)a^2>.
\end{equation}
In \citet{zhukovska2016} this quantity, and thus $\tau_{\rm acc}$, is constant for a given set of physical conditions. {We use the initial value of $\tau_{\rm acc}$ and Equation \ref{eq:gr2} to track the depletion in the case of a constant size distribution.}

\citet{zhukovska2016} implement dust destruction via two mechanisms: direct destruction in gas particles affected by supernovae in the simulation, using a prescription for the mass of gas `cleared' of dust \citep{jones1994,jones1996,dwek2007}, and additional destruction in the diffuse ISM representing supernovae originating from field OB stars, rather than those near their birth molecular clouds. The latter is treated as a destruction timescale $\tau_{\rm dest}$ in gas below a density threshold of $\nh = 1 \pcc$, with \citet{zhukovska2016} choosing $\tau_{\rm dest} = 100 \myr$. { From the evolution of the total destruction timescale presented in \citet{zhukovska2016}, we note that a) the second process appears to be dominant over destruction by individual supernova events, and b) $\tau_{\rm dest}$ remains between $\sim 300-400 \myr$, in approximate balance with the dust production rate, for the majority of the simulation.} We thus set $\tau_{\rm dest} = 350 \myr$, with the dust destruction then being given by
\begin{equation}
  \frac{df_d}{dt} = -\frac{f_d}{\tau_{\rm dest}}.
\end{equation}
This treatment does not account for grain size, with larger grains expected to be more resilient to destruction via sputtering. However, grain shattering processes, which are more efficient for larger grains, can redistribute the mass to smaller grain sizes, so that the post-shock size distribution is in general a complex function of many input parameters and assumptions \citep{slavin2015,kirchschlager2019}. As a full treatment of dust destruction is beyond the scope of this paper, we assume all grain sizes are affected equally, i.e. with the same $\tau_{\rm dest}$.

We initially investigate dust growth under typical CNM conditions of $\nh = 30 \pcc$ and $T = 100 \kel$ \citep{weingartner1999}. {We follow the evolution for $10 \myr$, typical of grain residence times in the CNM \citep{peters2017}, with a timestep of $10^4 \yr$.} Model parameters are listed in Table \ref{tab:params}.

\section{Results}

\begin{figure*}
  \centering
  \subfigure{\includegraphics[width=\columnwidth]{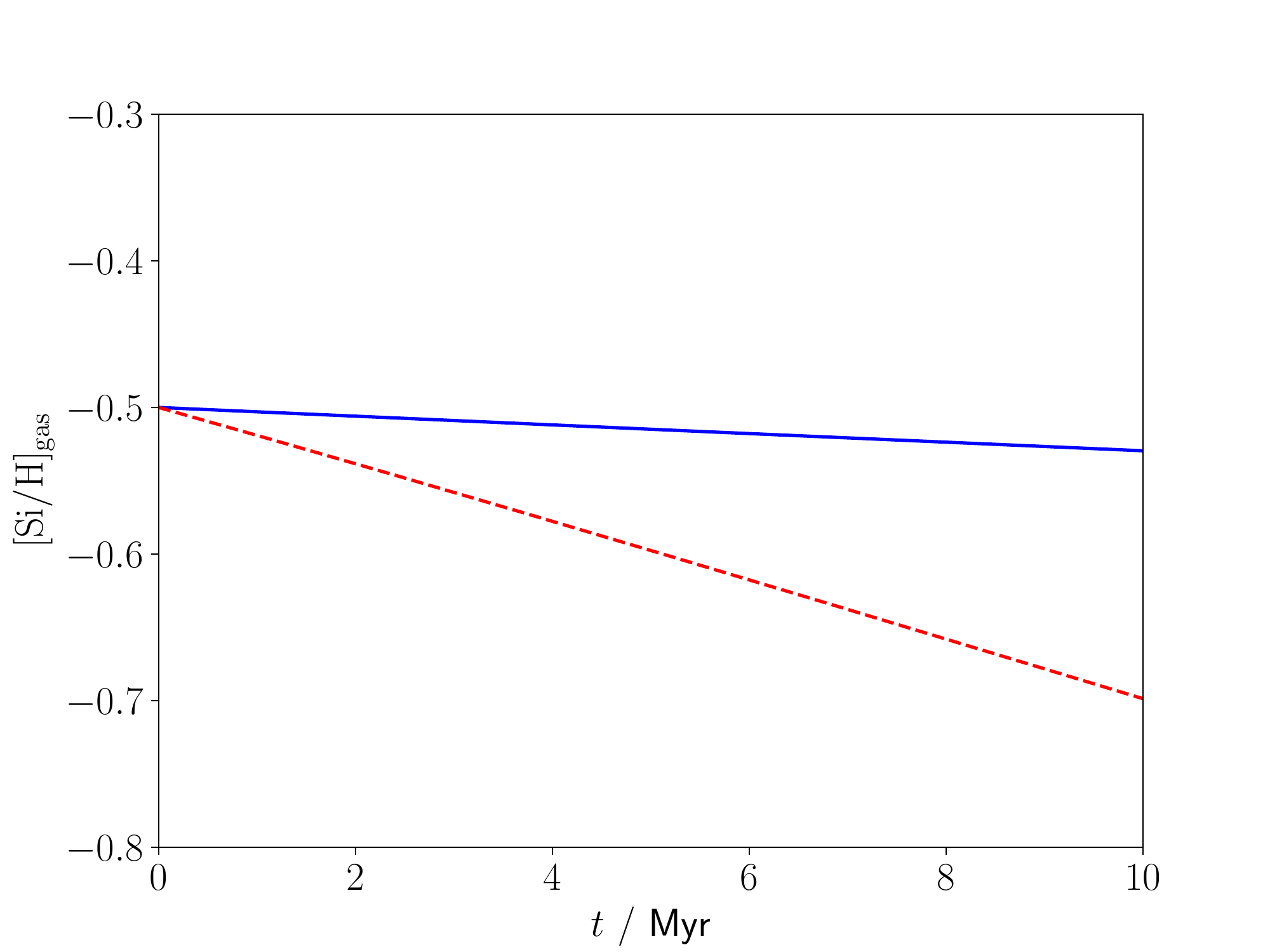}}\quad
  \subfigure{\includegraphics[width=\columnwidth]{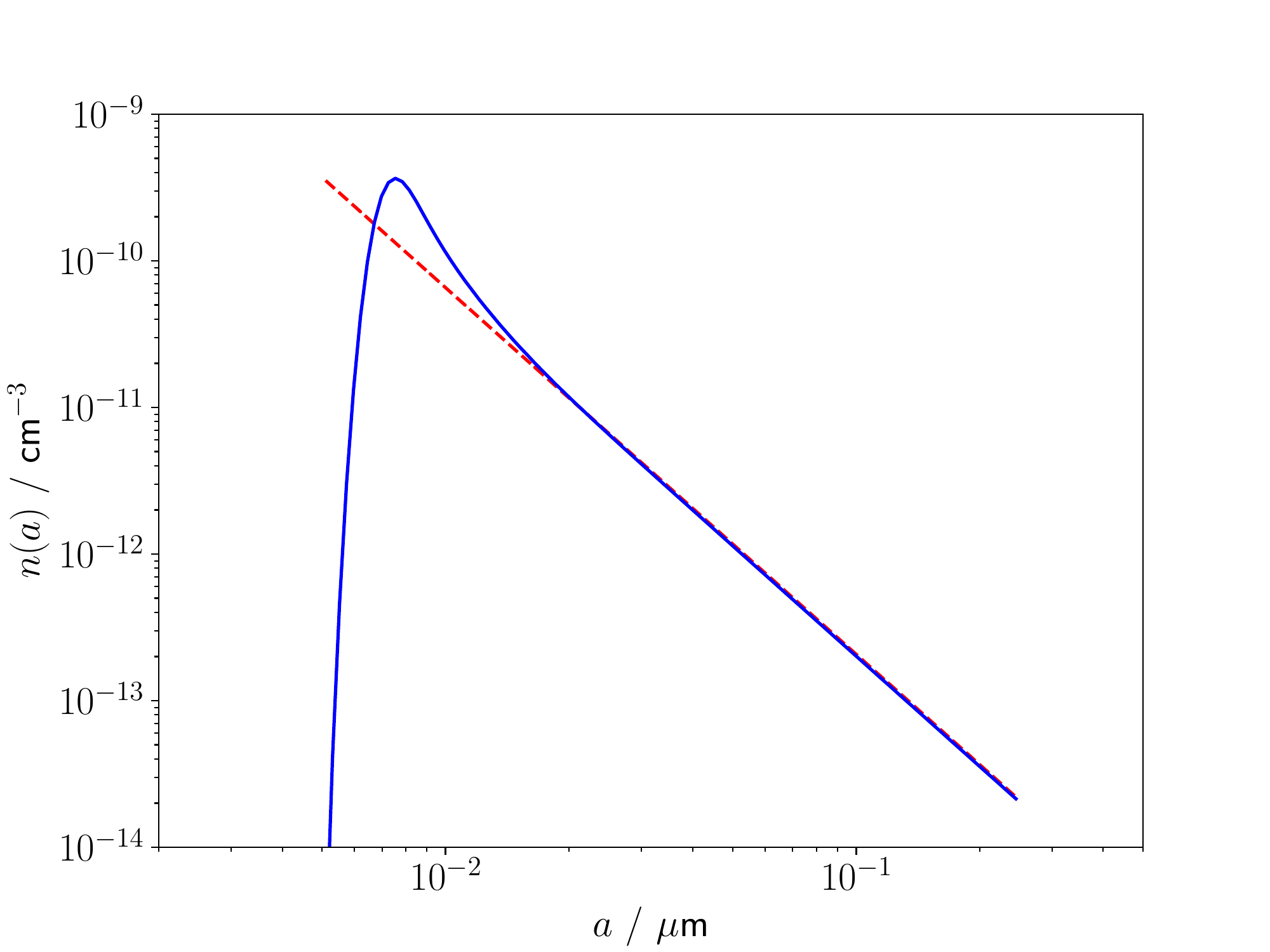}}
  \caption{Silicon depletion (left) and final grain size distribution (right) for models with an evolving size distribution (blue solid lines) or for a constant MRN distribution (red dashed lines), for $\nh = 30 \pcc$ and $T = 100 \kel$.}
  \label{fig:n30t100}
\end{figure*}

\begin{figure}
  \centering
  \includegraphics[width=\columnwidth]{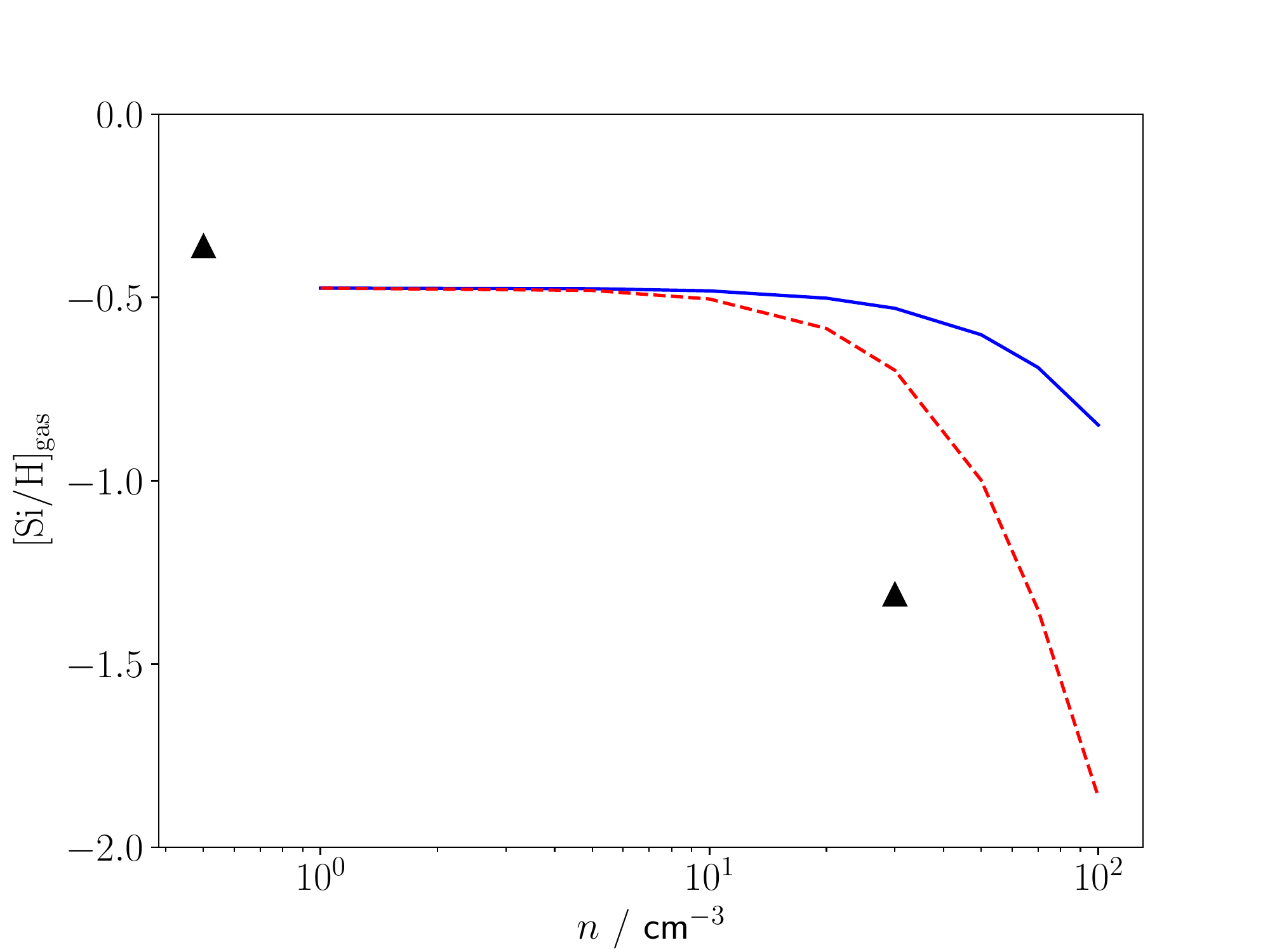}
  \caption{Silicon depletion after $10 \myr$ versus gas density for models with an evolving size distribution (blue solid line) or for a constant MRN distribution (red dashed line). Observational values from \citet{savage1996} for the warm and cool disc are shown as black triangles at representative densities of $\nh = 0.5$ and $30 \pcc$.}
  \label{fig:obsdep}
\end{figure}

\begin{figure*}
  \centering
  \subfigure{\includegraphics[width=\columnwidth]{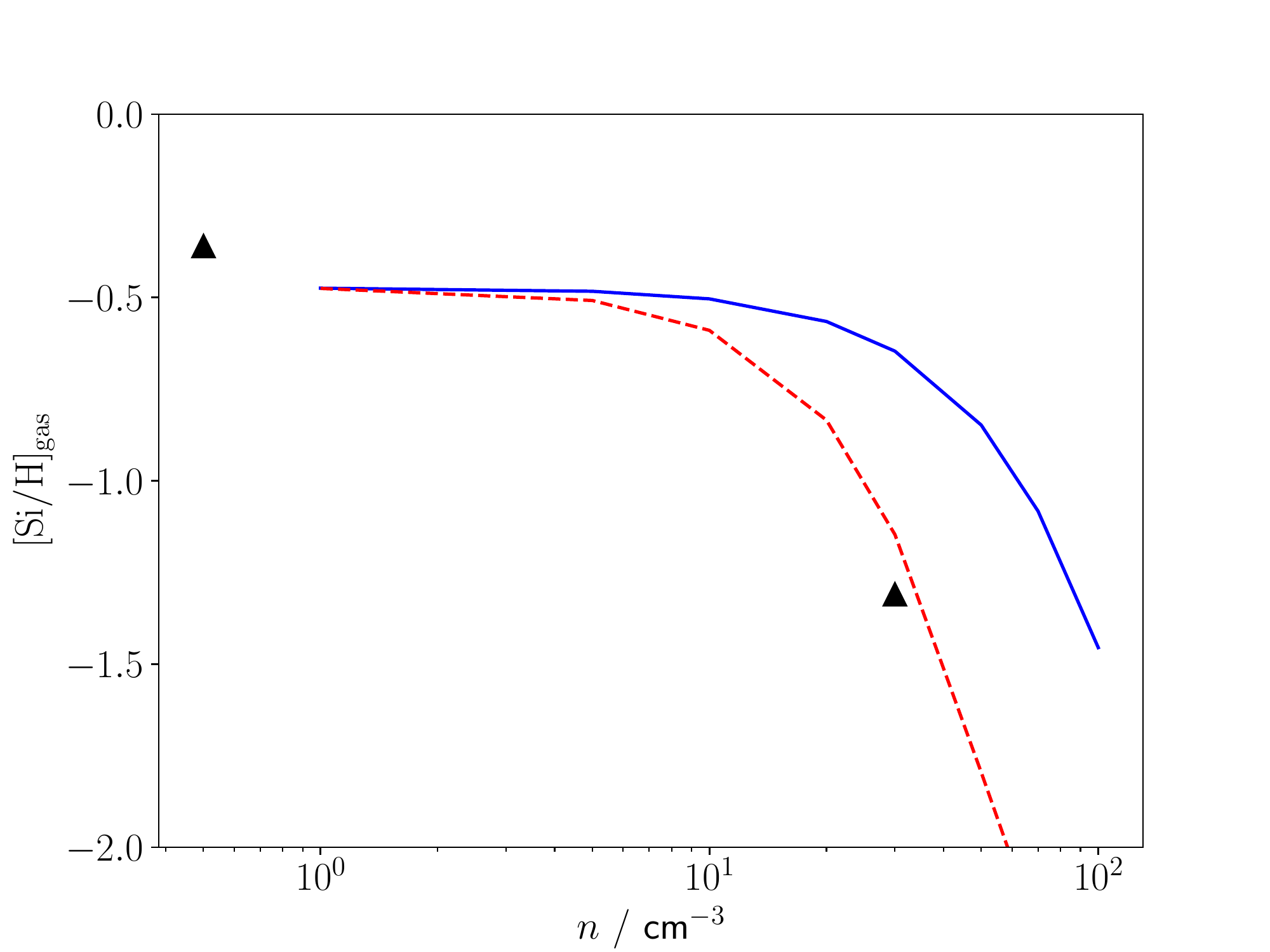}}\quad
  \subfigure{\includegraphics[width=\columnwidth]{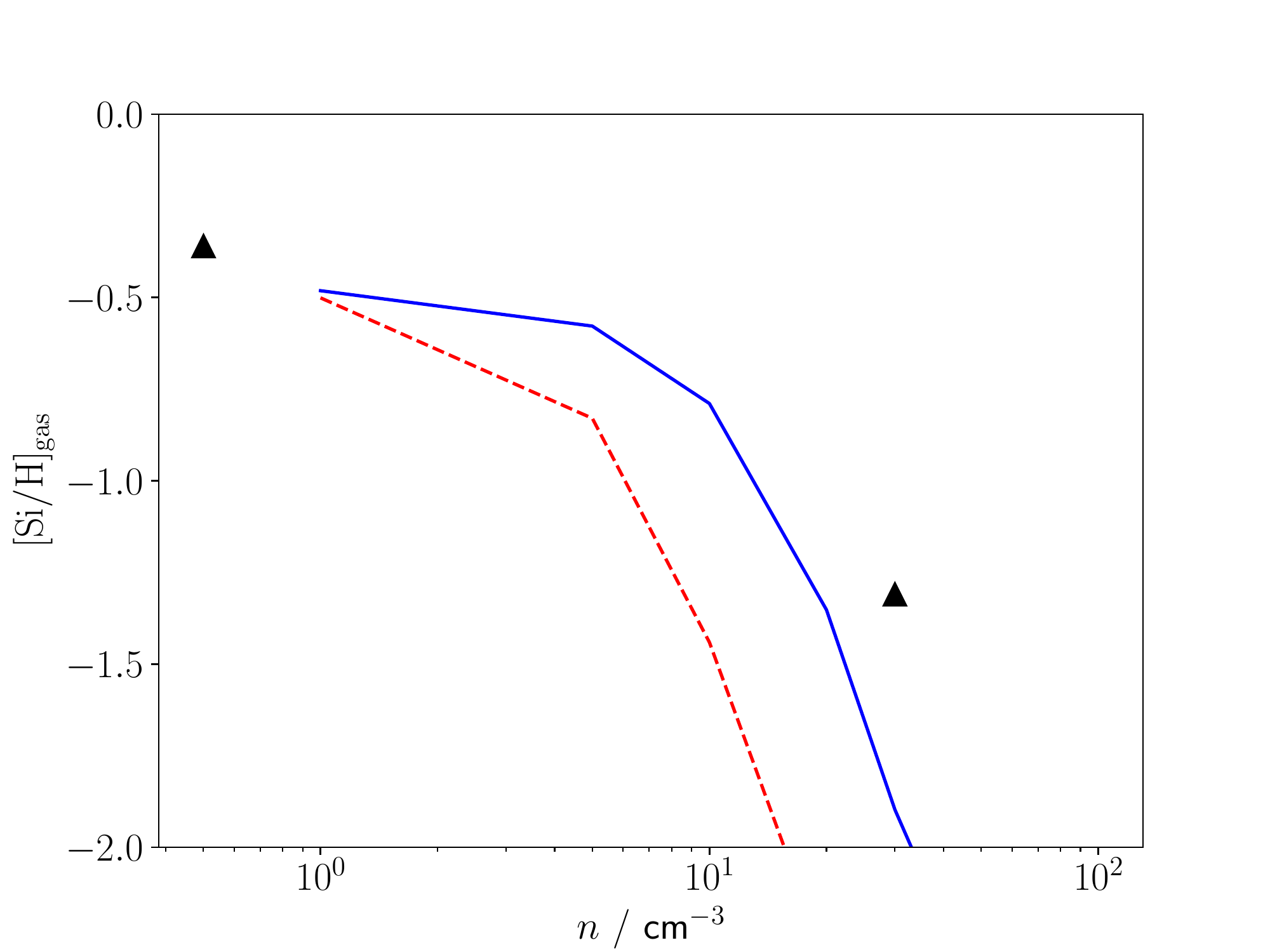}}
  \caption{Silicon depletion after $10 \myr$ versus gas density for models with an evolving size distribution (blue solid line) or for a constant MRN distribution (red dashed line), with $\amin = 3$ (left) and $1 \nm$ (right). Observational values from \citet{savage1996} for the warm and cool disc are shown as black triangles at representative densities of $\nh = 0.5$ and $30 \pcc$.}
  \label{fig:obsdepamin}
\end{figure*}

\begin{figure}
  \centering
  \includegraphics[width=\columnwidth]{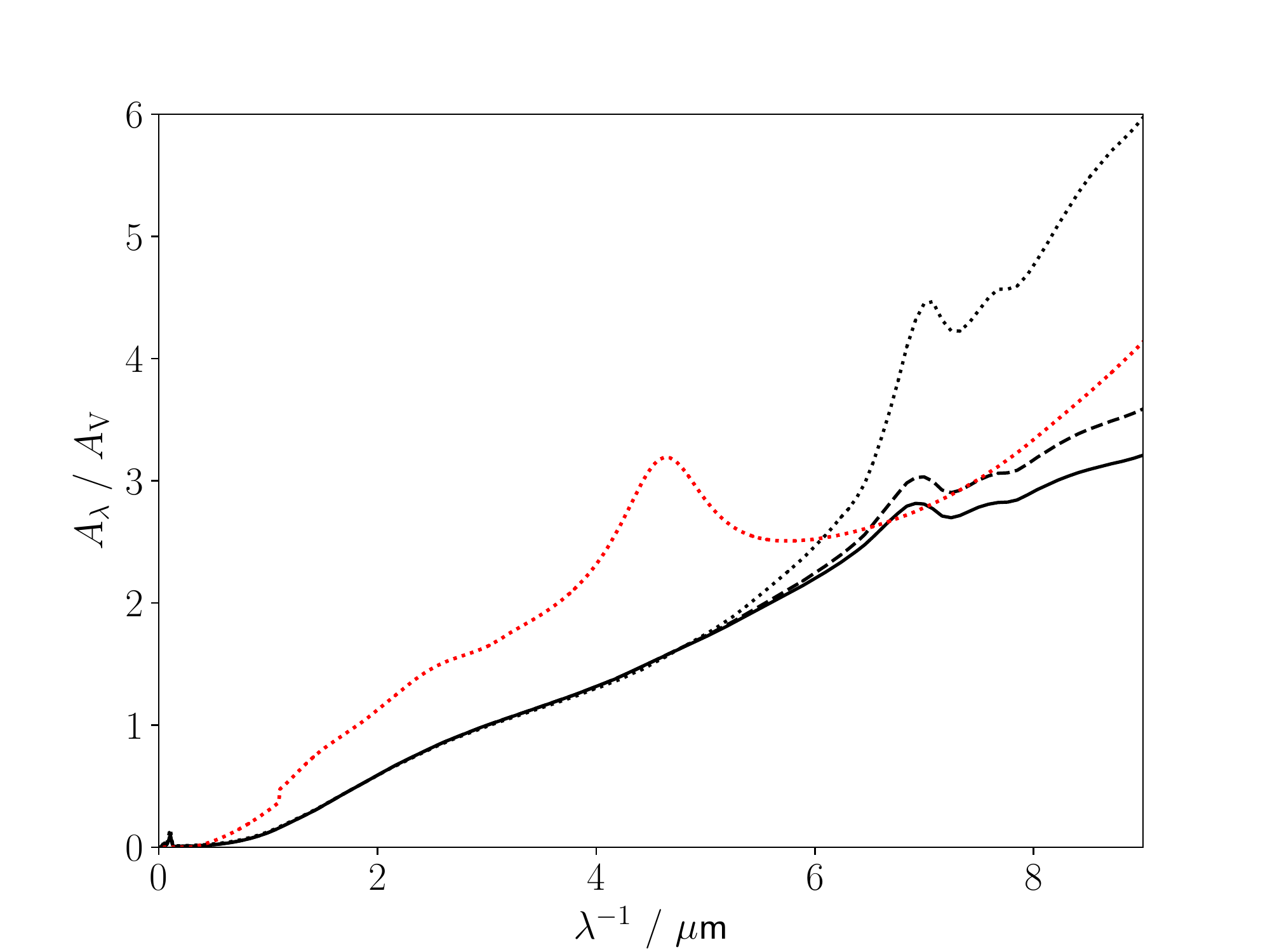}
  \caption{Extinction curves, using \citet{laor1993} silicate optical properties, for the initial MRN distribution (solid black line) and the grain size distributions after $10 \myr$ for our fiducial model, with $\amin = 5 \nm$ (dashed black line), and for $\amin = 1 \nm$ (dotted black line). {The values of $A_{\rm V}/N_{\rm H}$ are $1.2$, $1.2$ and $1.1 \times 10^{-22} \, {\rm mag \, cm^2}$ respectively. The value of $A_{\lambda}/A_{\rm V}$ has been reduced by half to account for the typical silicate/carbon ratio of grains in the ISM.} The \citet{cardelli1989} Galactic extinction curve, with $R_{\rm V} = 3.1$, is also shown for comparison (red dotted line).}
  \label{fig:sizeext}
\end{figure}

\begin{figure}
  \centering
  \subfigure{\includegraphics[width=\columnwidth]{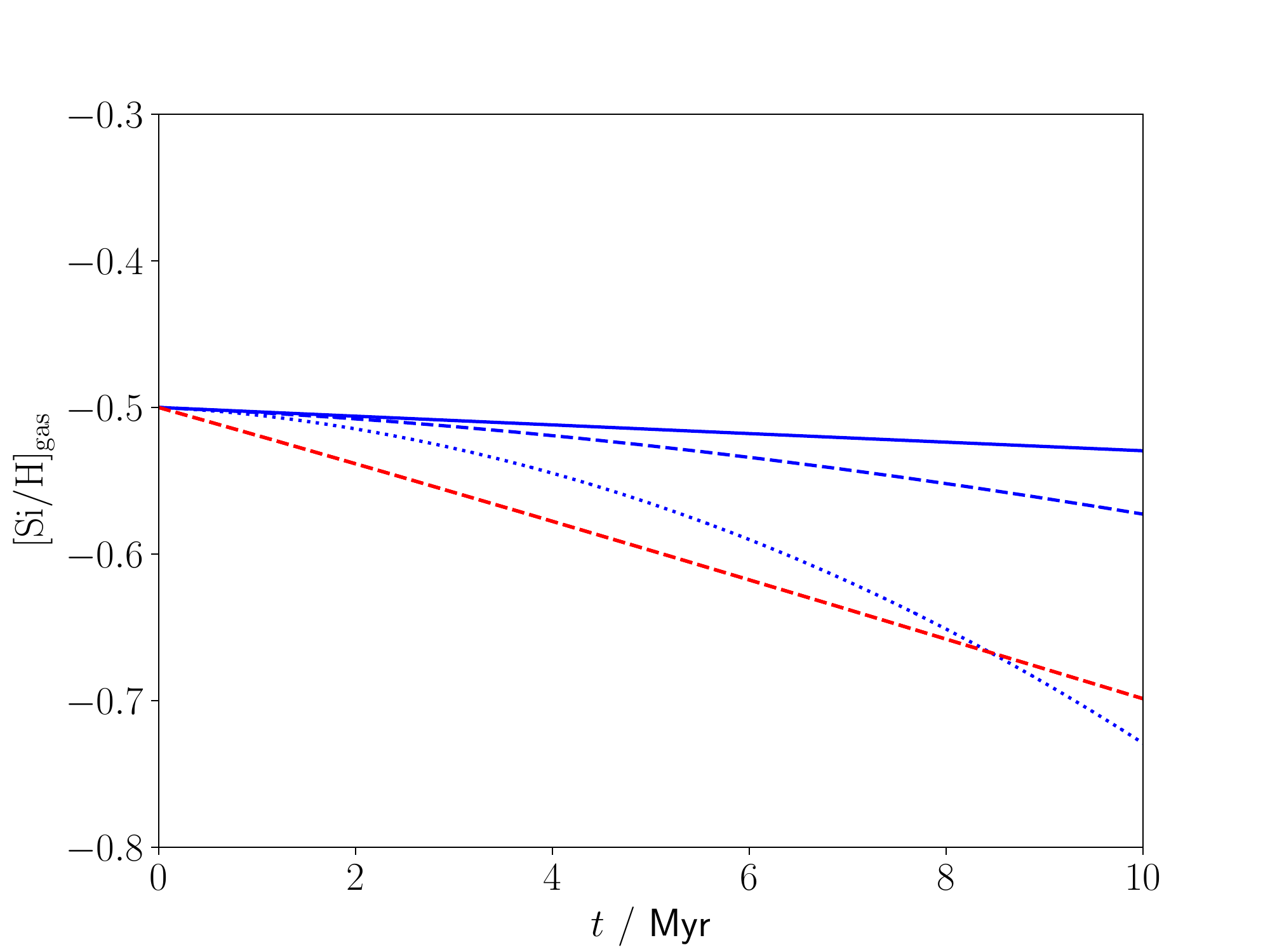}}
  \caption{Silicon depletion for models with an evolving size distribution (blue lines) or for a constant MRN distribution (red dashed line), for $\nh = 30 \pcc$ and $T = 100 \kel$, and a shattering efficiency of $0$ (solid blue line), $0.01$ (dashed blue line) and $0.05$ (dotted blue line).}
  \label{fig:shatter}
\end{figure}

\begin{figure*}
  \centering
  \subfigure{\includegraphics[width=\columnwidth]{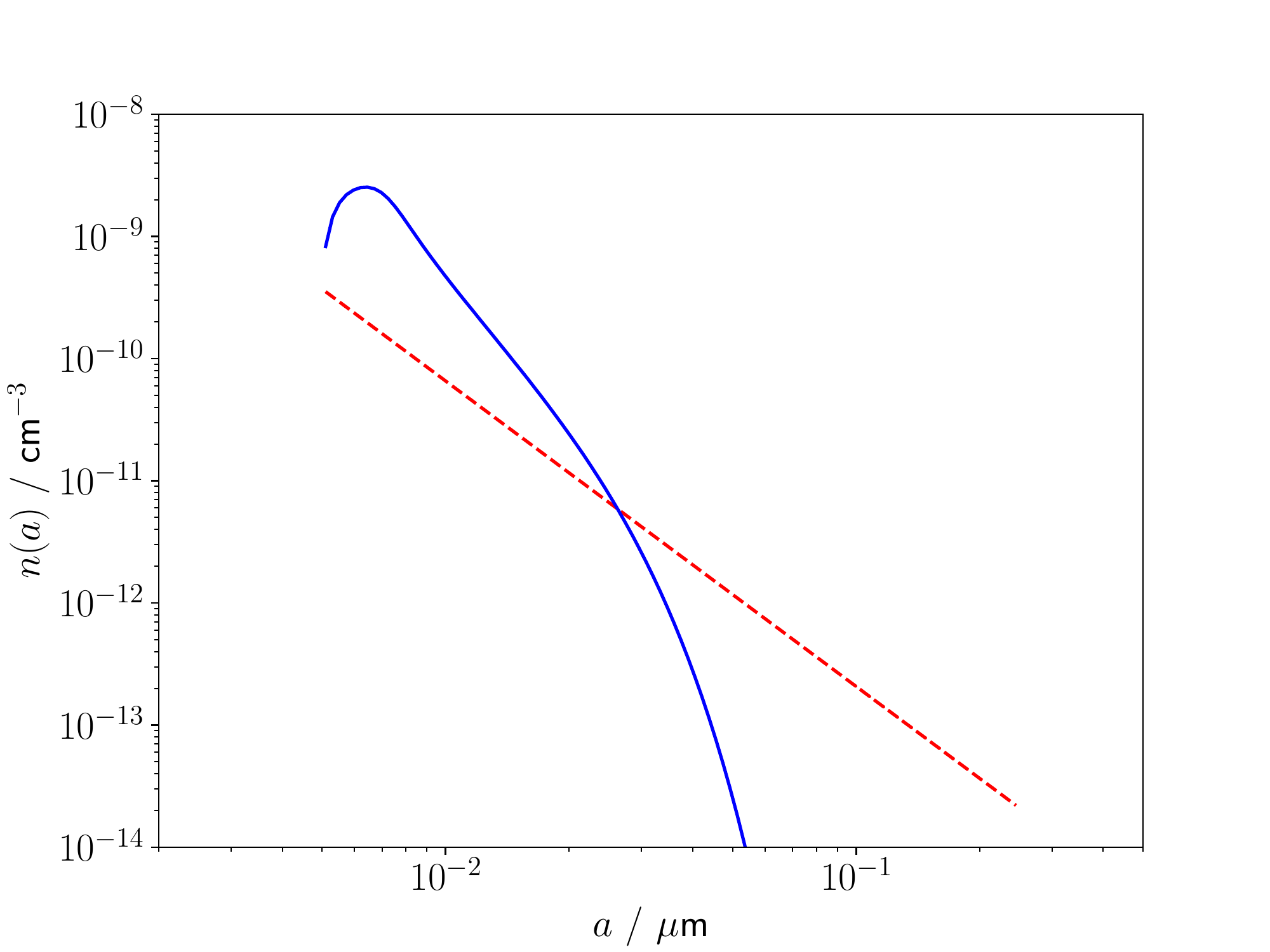}}\quad
  \subfigure{\includegraphics[width=\columnwidth]{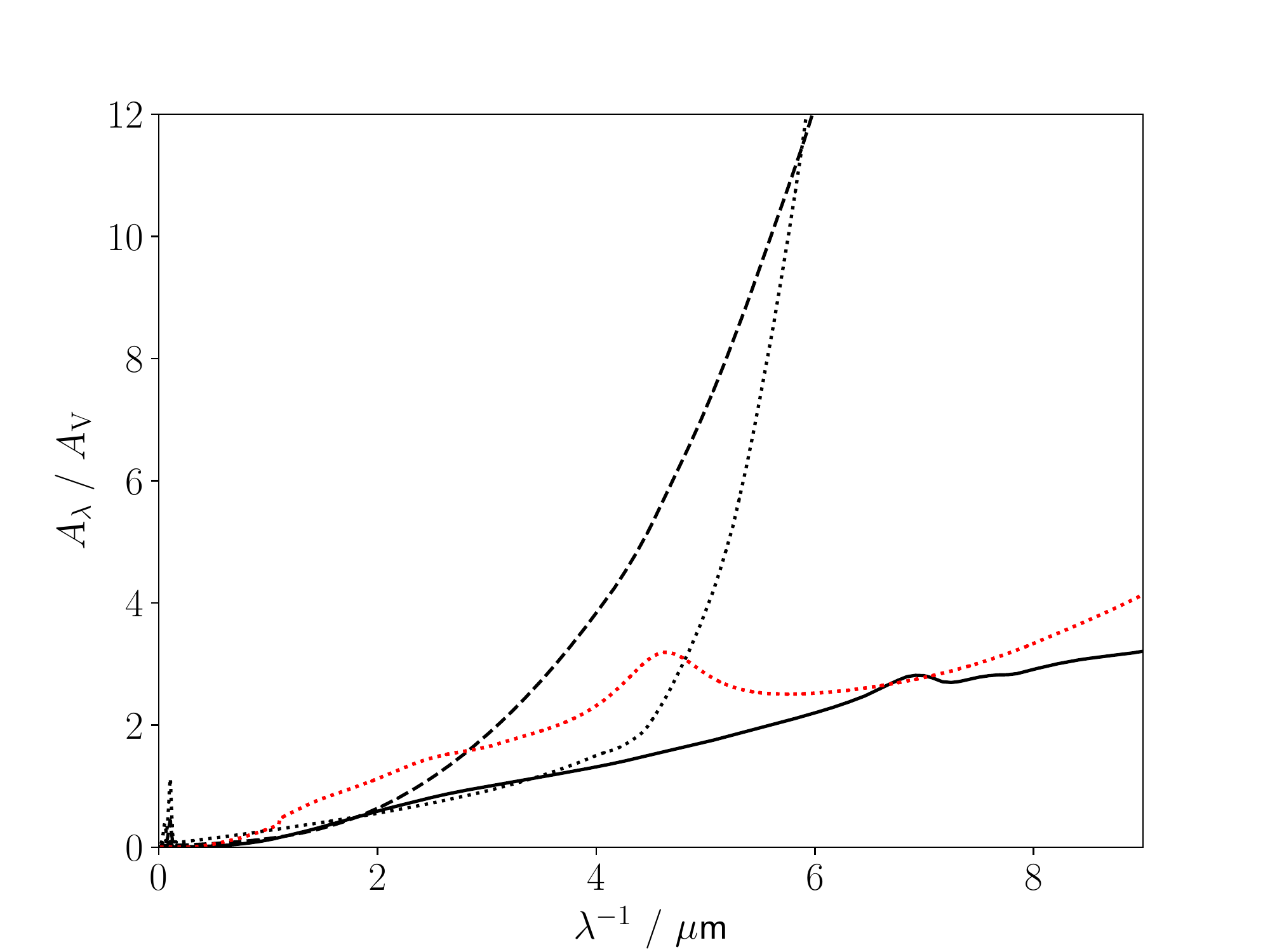}}
  \caption{\textit{Left:} Final grain size distribution (blue solid line) and initial MRN distribution (red dashed line) after $10 \myr$ for $\nh = 30 \pcc$, $T = 100 \kel$, and a shattering efficiency of $0.05$. \textit{Right:} Extinction curves, using \citet{laor1993} silicate optical properties, for the initial MRN distribution (solid black line) and the grain size distributions after $10 \myr$ for a shattering efficiency of $0.01$ (dashed black line) and $0.05$ (dotted black line). {The values of $A_{\rm V}/N_{\rm H}$ are $1.2$, $0.28$ and $0.12 \times 10^{-22} \, {\rm mag \, cm^2}$ respectively. The value of $A_{\lambda}/A_{\rm V}$ has been reduced by half to account for the typical silicate/carbon ratio of grains in the ISM.} The \citet{cardelli1989} Galactic extinction curve, with $R_{\rm V} = 3.1$, is also shown for comparison (red dotted line).}
  \label{fig:shatterext}
\end{figure*}

\begin{figure*}
  \centering
  \subfigure{\includegraphics[width=\columnwidth]{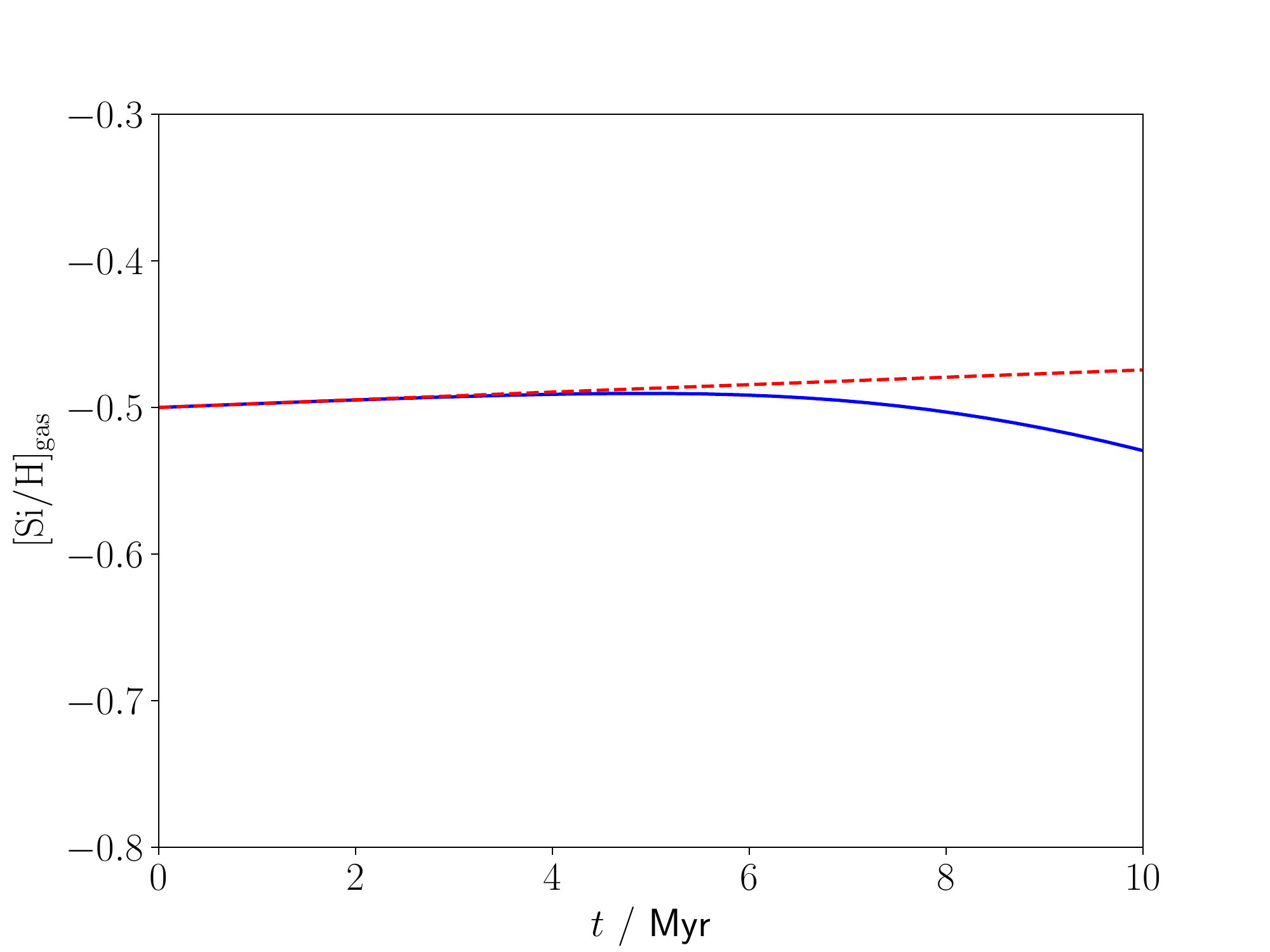}}\quad
  \subfigure{\includegraphics[width=\columnwidth]{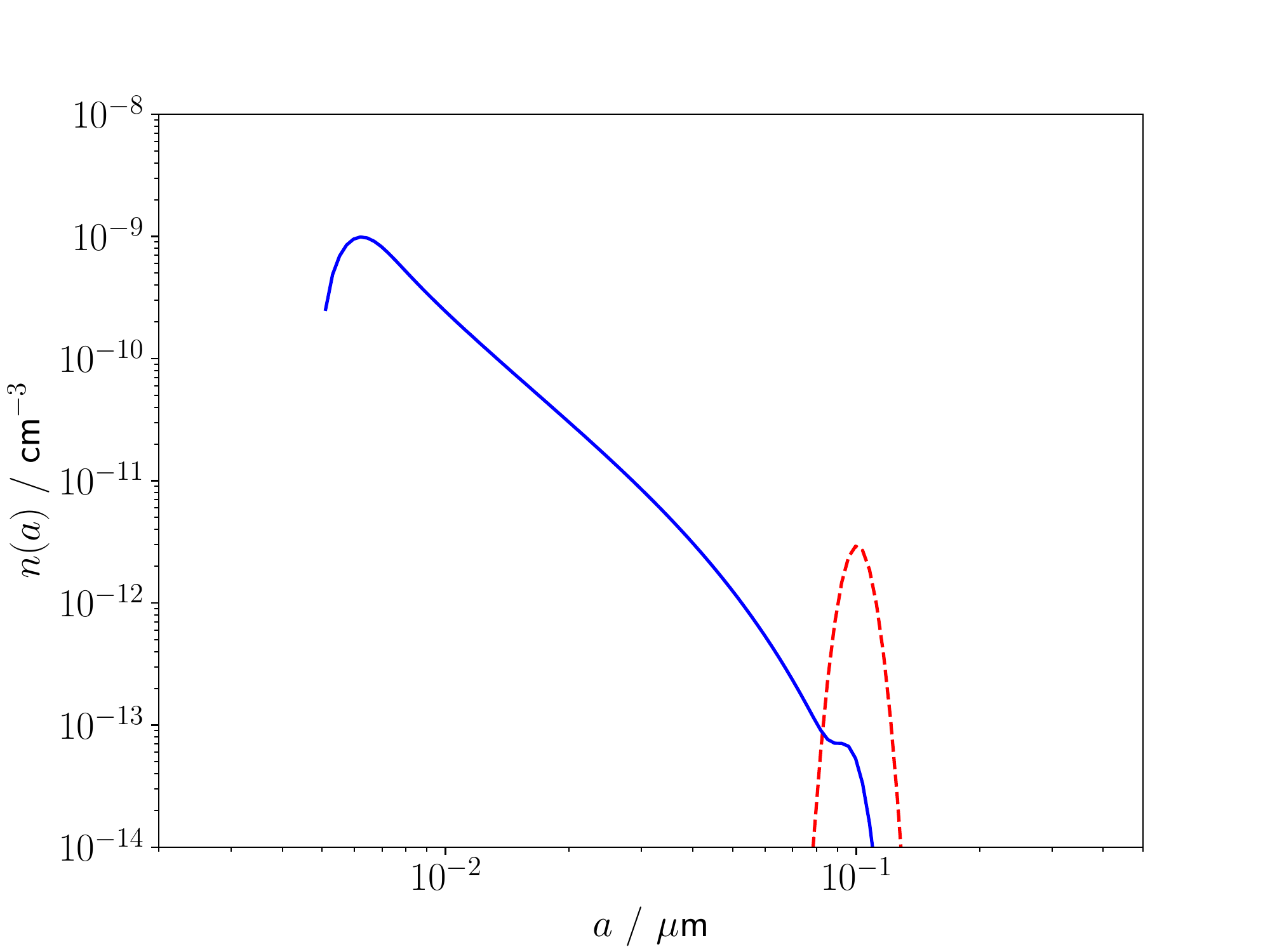}}
  \caption{Silicon depletion (left) and final grain size distribution (right) for models with an evolving (blue solid lines) and constant size distribution (red dashed lines), for $\nh = 30 \pcc$, $T = 100 \kel$ and a shattering efficiency of $0.05$, and an initial log-normal size distribution centred at and with a width of $0.1 \um$.}
  \label{fig:bigdust}
\end{figure*}

{Figure \ref{fig:n30t100} shows the evolution of $\depsi$ and the grain size distribution for $\nh = 30 \pcc$, $T = 100 \kel$, with and without an evolving grain size distribution. For a constant MRN distribution, the level of silicon depletion increases by $0.2$ dex over $10 \myr$, whereas when the increase in grain radii is accounted for the increase is negligible. The final size distribution develops a peak at $\sim 0.01 \um$, where the typical grain charge starts to transition from negative to positive. Grain smaller than this can accrete gas-phase material efficiently, but at larger radii this is increasingly prevented as the Coulomb focusing factor decreases, and large, positively-charged grains are almost completely unable to grow via accretion. This effect occurs regardless of gas properties - Figure \ref{fig:obsdep} shows $\depsi$ after $10 \myr$ for varying gas density, compared to representative observational values taken from \citet{savage1996}. The increase in the size of the smallest grains, and the corresponding reduction in the accretion efficiency, results in less depletion at all densities, but this is particularly noticeable at the higher densities, where the timescales are short enough for significant accretion to occur over the $10 \myr$ timespan.}

{Even for a constant size distribution, our model returns values of $\depsi$ significantly higher than those observed in the CNM. \citet{zhukovska2016} found the same issue for the typical MRN minimum grain size of $5 \nm$, motivating them to investigate smaller values. Figure \ref{fig:obsdepamin} shows $\depsi$ versus gas density for minimum grain radii of $3$ and $1 \nm$. With $\amin = 3 \nm$ and a constant MRN distribution, the \citet{savage1996} CNM depletion can be reproduced, as found by \citet{zhukovska2016}, but with an evolving size distribution this value ($-1.3$ dex) is only reached for a gas density of $\nh = 100 \pcc$, at which point several of our model assumptions, such as the temperature and electron density, are likely to become inappropriate. Reducing the minimum grain size further to $1 \nm$ allows $\depsi$ to reach the observed value within $10 \myr$, but the resulting size distribution is significantly overabundant in $\sim 0.01 \um$ grains compared to the typical MRN case. This causes a corresponding increase in the far-UV extinction which is incompatible with values seen along Galactic sightlines, shown in Figure \ref{fig:sizeext} (we have reduced the value of $A_{\lambda}/A_{\rm V}$ by half to account for the fact that silicates make up only $\sim 50\%$ of the ISM dust budget).} {The model ratios of visual extinction to column density, $A_{\rm V}/N_{\rm H}$, are $\sim 10^{-22} \, {\rm mag \, cm^2}$, lower than expected compared to Galactic values ($\sim 5 \times 10^{-22} \, {\rm mag \, cm^2}$; \citealt{bohlin1978}) if silicates make up half of the total $A_{\rm V}$. This suggests that large grains are underabundant, even for the initial size distribution, due to the low initial depletion and lack of a mechanism to produce grains larger than $\sim 0.01 \um$.}

{The reduction in accretion efficiency is caused by the bottlenecking of small grains as they reach radii where they become positively charged, at which point further growth becomes very slow. In principle, this could be mitigated by grain shattering, redistributing mass from large grains into more numerous, smaller ones. However, we find this either fails to resolve the issue, or raises new ones. A full treatment of shattering requires knowledge of both the dynamics and bulk properties of the dust grains \citep{hirashita2009,kirchschlager2019}, and is beyond the scope of this paper, but we can approximate its effect by calculating the rate of grain-grain collisions and assuming that a given fraction (hereafter the `shattering efficiency') result in a shattering event. Following \citet{hirashita2009}, we assume a typical turbulent velocity dispersion in the CNM of $2 \kms$, and that the mass of shattered grains is redistributed into a MRN power law size distribution, with the maximum mass equal to the size of the shattered grain.}

{Figure \ref{fig:shatter} shows the evolution of $\depsi$ for models with varying shattering efficiency. Reproducing the static size distribution values requires a shattering efficiency of $\gtrsim 0.01$, but the size distributions and extinction curves produced with these values, shown in Figure \ref{fig:shatterext}, are in even greater tension with those observed than the reduced $\amin$ models discussed above - {the extinction in the far-UV is even higher, and $A_{\rm V}/N_{\rm H}$ even lower, due to the rapid redistribution of mass from large to small grains}. We also consider this level of shattering to be unrealistic - the threshold velocity for shattering of silicates in \citet{hirashita2009} is $2.7 \kms$, larger than the CNM turbulent velocity, so the number of collisions resulting in a shattering event is likely to be very small. The CNM size distribution presented in \citet{hirashita2009} is almost unchanged after $50 \myr$ of evolution, and does not display the large overabundance of small grains required for efficient grain growth.}

{The inability of grain growth models to reproduce the observed elemental depletions without also producing unrealistic extinction curves is tied to the initial size distribution - the amount of possible growth is limited by the number of small grains, which can only accrete up to a certain radius before becoming positively charged, and for an MRN distribution the amount of additional growth required to reach $\depsi \sim -1.3$ (as required by \citealt{savage1996}) results in implausible levels of far-UV extinction. It is possible that these issues could be circumvented by an initial size distribution with fewer small grains, and subsequent reprocessing of larger grains into smaller ones via shattering. However, we are unable to find a situation where this occurs. Figure \ref{fig:bigdust} shows the evolution of $\depsi$ and the grain size distribution for the extreme case of a top-heavy size distribution (a log-normal with centre and width $0.1 \um$). Such a size distribution could plausibly be produced by the injection of dust by core-collapse supernovae, which are established as primarily producing large grains \citep{gall2014,wesson2015,bevan2016,priestley2020}; from hotter phases of the ISM where only the largest grains are expected to survive; or from coagulation in molecular clouds. Even with a shattering efficiency of $0.05$, accretion of gas-phase material is still not efficient enough to reach observed depletion levels in the CNM after $10 \myr$, and the final size distribution is skewed far enough to small grain sizes to result in unrealistic levels of far-UV extinction. It may be possible to reproduce the \citet{savage1996} CNM depletions via grain growth without violating constraints on the far-UV extinction with a suitable choice of initial conditions, but we are unable to find a physically motivated scenario for which this is the case.}

\section{Discussion}

\begin{figure}
  \centering
  \includegraphics[width=\columnwidth]{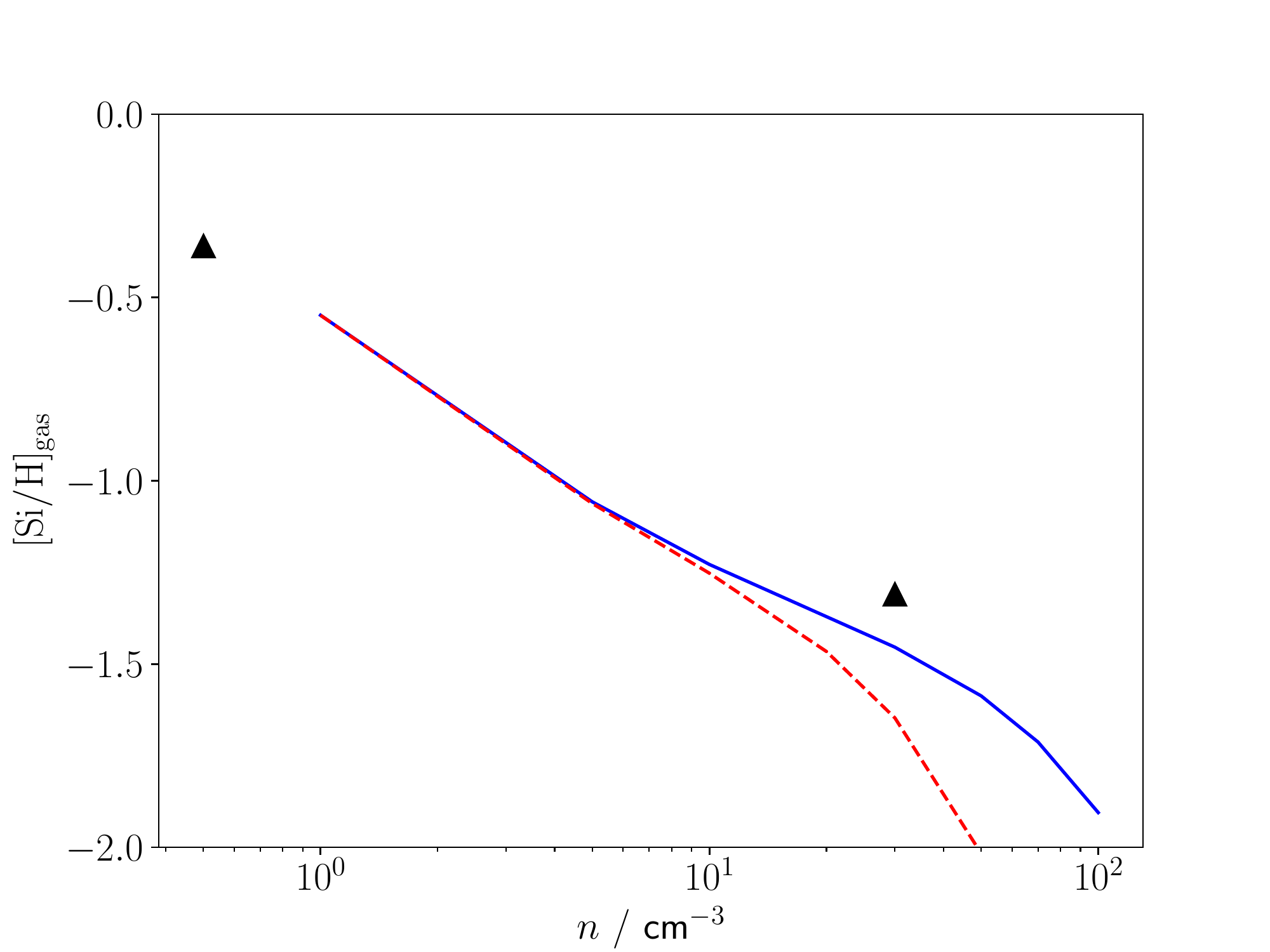}
  \caption{Silicon depletion after $10 \myr$ versus gas density for models with an evolving size distribution (blue solid line) or for a constant MRN distribution (red dashed line), assuming an initial depletion of $\depsi = -1.5$ and a dust destruction timescale $\tau_{\rm dest} = 1 {\rm \, Gyr} \, (\nh/30 \pcc)$. Observational values from \citet{savage1996} for the warm and cool disc are shown as black triangles at representative densities of $\nh = 0.5$ and $30 \pcc$.}
  \label{fig:lowdep}
\end{figure}

\begin{figure*}
  \centering
  \subfigure{\includegraphics[width=\columnwidth]{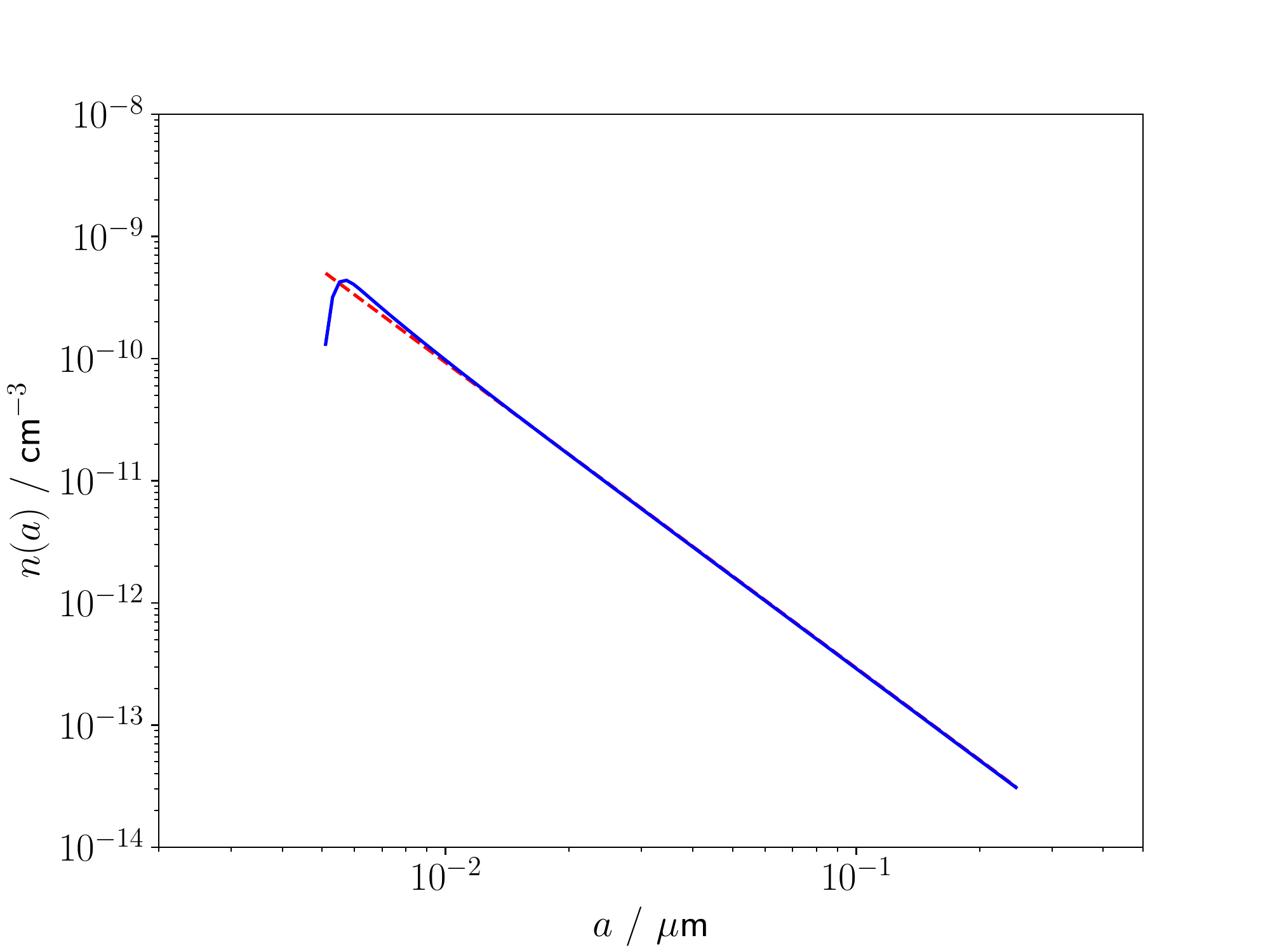}}\quad
  \subfigure{\includegraphics[width=\columnwidth]{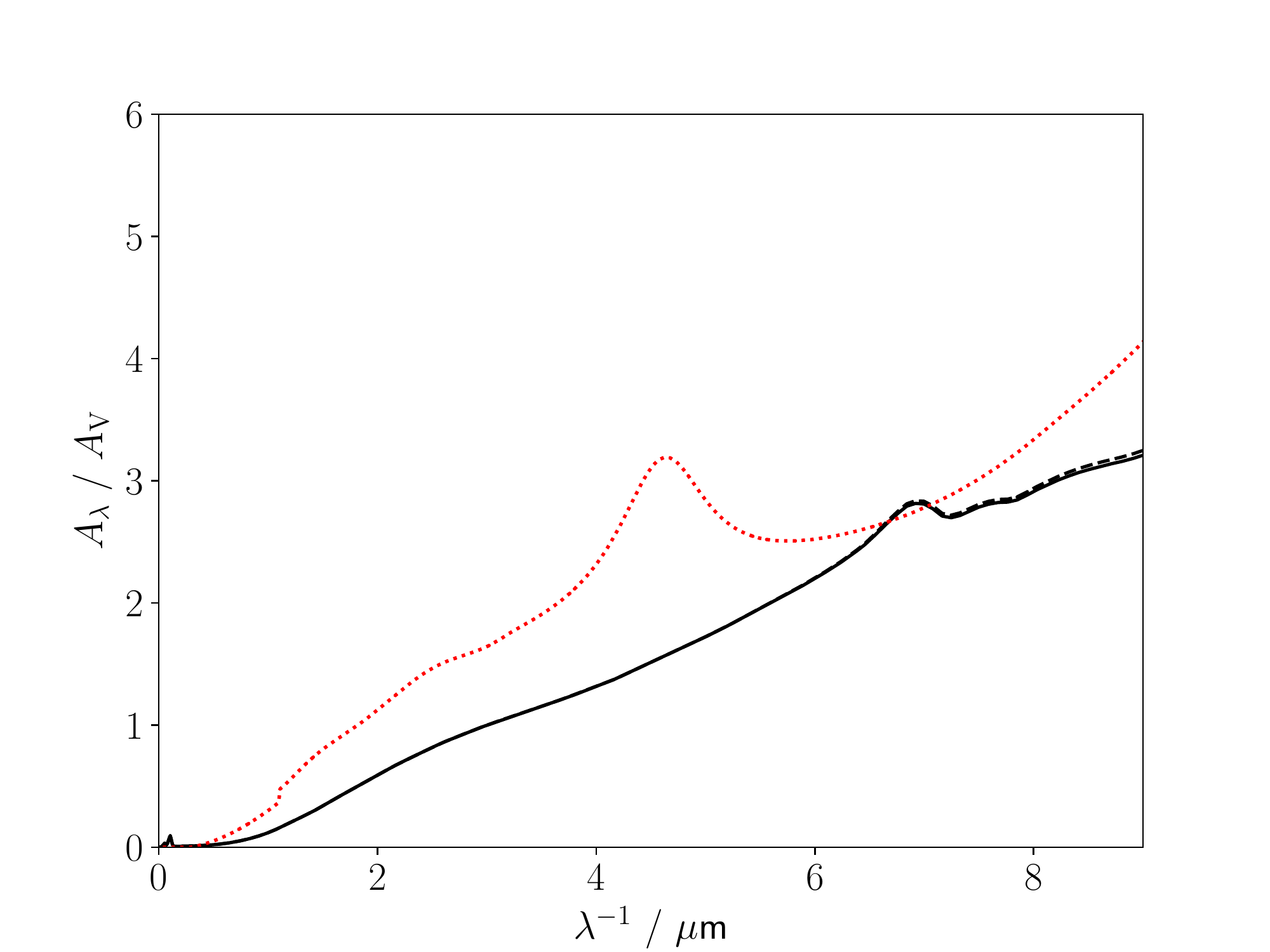}}
  \caption{\textit{Left:} Final grain size distribution (blue solid line) and initial MRN distribution (red dashed line) after $10 \myr$ for $\nh = 30 \pcc$, $T = 100 \kel$, {$\depsi = -1.5$ and  $\tau_{\rm dest} = 1 \, {\rm Gyr}$}. \textit{Right:} Extinction curves, using \citet{laor1993} silicate optical properties, for the initial (solid black line) {and final (dashed black line) size distributions}. {The values of $A_{\rm V}/N_{\rm H}$ are $1.7$ and $1.7 \times 10^{-22} \, {\rm mag \, cm^2}$ respectively.} The value of $A_{\lambda}/A_{\rm V}$ has been reduced by half to account for the typical silicate/carbon ratio of grains in the ISM. The \citet{cardelli1989} Galactic extinction curve, with $R_{\rm V} = 3.1$, is also shown for comparison (red dotted line).}
  \label{fig:lowdepext}
\end{figure*}

{If observed elemental depletions in the CNM cannot be reproduced by efficient accretion onto small grains, they must be caused by some other mechanism. We suggest that the assumed dust destruction timescales are the most likely avenue. As noted above, \citet{zhukovska2016} account for both destruction by supernovae captured in their hydrodynamical simulations, and additional destruction in diffuse gas ascribed to supernovae occuring at a distance from their birth molecular clouds. The latter mechanism is introduced in order to prevent overdepletion in the CNM, which may not be necessary if the assumed grain-growth timescales are too low as we argue, and the former is based on one-dimensional theoretical models which appear to be in conflict with at least some observed supernova remnants \citep{koo2016,chawner2020b,priestley2021}. If either or both of the assumed destruction timescales are too low, the low gas-phase silicon abundance in the CNM is not due to efficient accretion of gas-phase material, but is indicative of highly-depleted gas injected into this phase of the ISM. Observations of both the Crab Nebula \citep{owen2015,delooze2019} and Cassiopeia A \citep{delooze2017,laming2020} have found that supernova ejecta dust masses are comparable to those of the gas-phase dust-forming elements, or possibly even higher. High initial depletions could also be due to grain accretion in molecular clouds, although this would require some way of growing the refractory mass rather than ice mantles.}

{Figure \ref{fig:lowdep} demonstrates an attempt at constructing a model based on these hypotheses. We assume that an initial value of $\depsi = -1.5$, and a dust destruction timescale scaling as $\tau_{\rm dest} = 1 {\rm \, Gyr} \, (\nh/30 \pcc)$. The normalisation of $\tau_{\rm dest}$ is motivated by the values in \citet{zhukovska2016} without the additional diffuse ISM destruction, while the scaling is consistent with previous theoretical work on the density dependence of dust destruction \citep{draine1990,hu2019}. With or without an evolving size distribution, the model broadly reproduces the depletions in both the CNM and warmer gas, as well as the trend of increasing depletion at higher densities. {The size distribution and extinction curve, shown in Figure \ref{fig:lowdepext}, are nearly unchanged from the initial conditions after $10 \myr$ at $\nh = 30 \pcc$, and with $A_{\rm V}/N_{\rm H} = 1.7 \times 10^{-22} \, {\rm mag \, cm^2}$ are in better agreement with \citet{bohlin1978}, particularly as we expect carbon grains to contribute slightly more to the total $A_{\rm V}$ (using amorphous carbon optical properties from \citet{zubko1996} instead of silicates, we find $A_{\rm V}/N_{\rm H} = 4.1 \times 10^{-22} \, {\rm mag \, cm^2}$).} While this is a somewhat artificial example due to the choice of the initial depletion, and neglects the cycling of gas between different phases of the ISM, it does show that phenomena commonly attributed to grain growth can be equally well explained by variations in the dust destruction efficiency.}

\section{Conclusions}

{Efficient accretion of gas-phase metals by small, negatively-charged dust grains has been proposed as an explanation for elemental depletion patterns observed in the CNM \citep{zhukovska2016,zhukovska2018}. We have demonstrated that once the evolution of the size distribution is properly accounted for, this becomes impossible, as the growth in dust mass is halted once the small grains grow large enough to become positively charged. Increasing or replenishing the number of small grains, such as by altering the initial size distribution or invoking efficient grain shattering, results in far-UV extinction curves incompatible with anything observed along Galactic sightlines. We suggest that relatively high levels of depletion in the CNM, rather than being a sign of efficient grain growth, are actually indicative of {the survival of dust grains in initially highly-depleted material from a (presumably) denser phase of the ISM.}}

\section*{Acknowledgements}

{We thank the referee for a useful report which significantly improved this paper.} We are grateful to Luisa Lucie-Smith for constructive discussions on this topic. FDP is funded by the Science and Technology Facilities Council. IDL acknowledges support from European Research Council (ERC) starting grant 851622 DustOrigin. MJB acknowledges support from the ERC grant SNDUST ERC-2015-AdG-694520.

\section*{Data Availability}

The data underlying this article will be made available upon request. The Fortran code used to generate the data is available at \url{www.github.com/fpriestley/growth}.




\bibliographystyle{mnras}
\bibliography{growth}


\appendix

\section{Reducing the CNM electron fraction}
\label{sec:lowe}

\begin{figure*}
  \centering
  \subfigure{\includegraphics[width=\columnwidth]{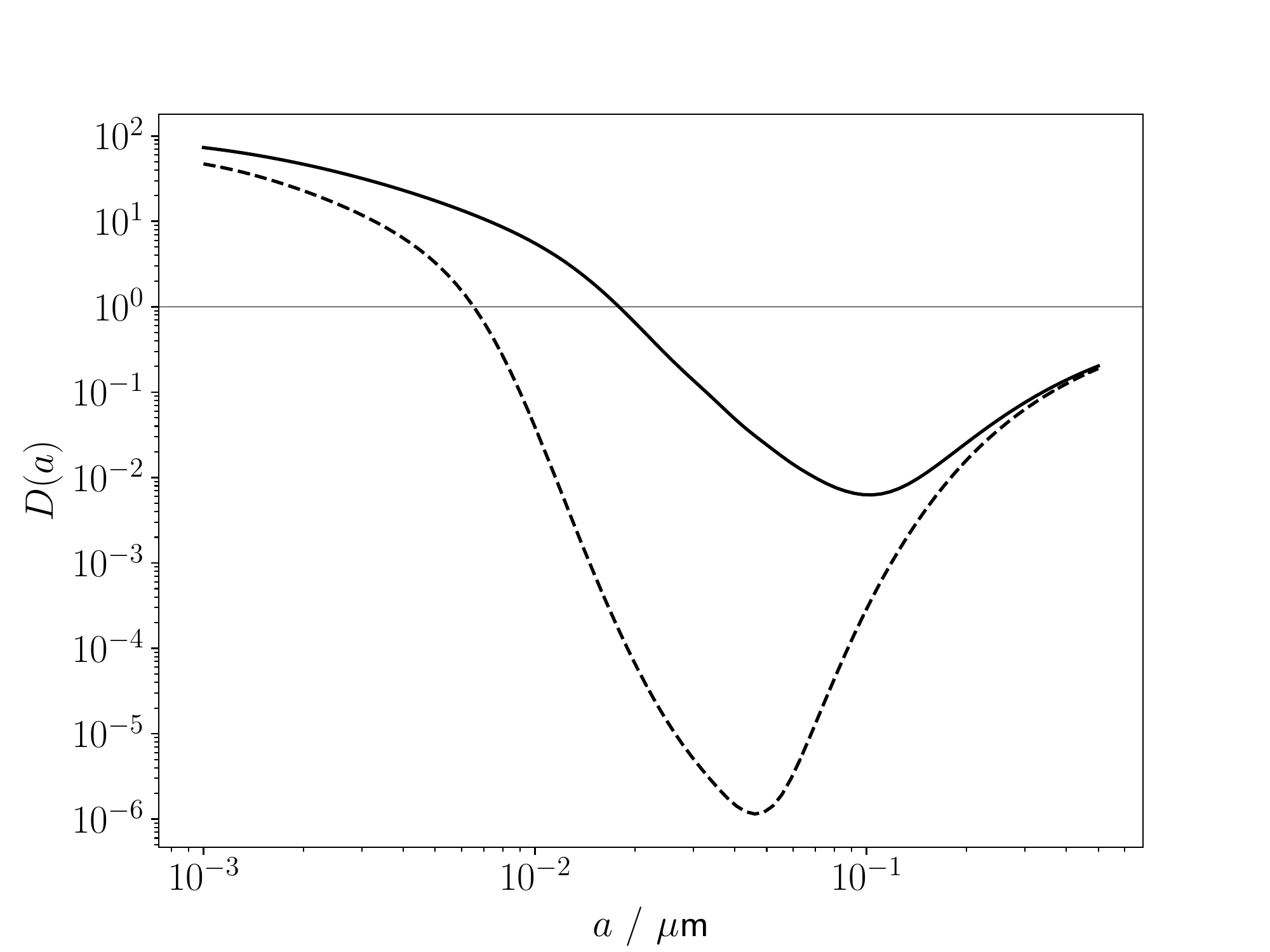}}\quad
  \subfigure{\includegraphics[width=\columnwidth]{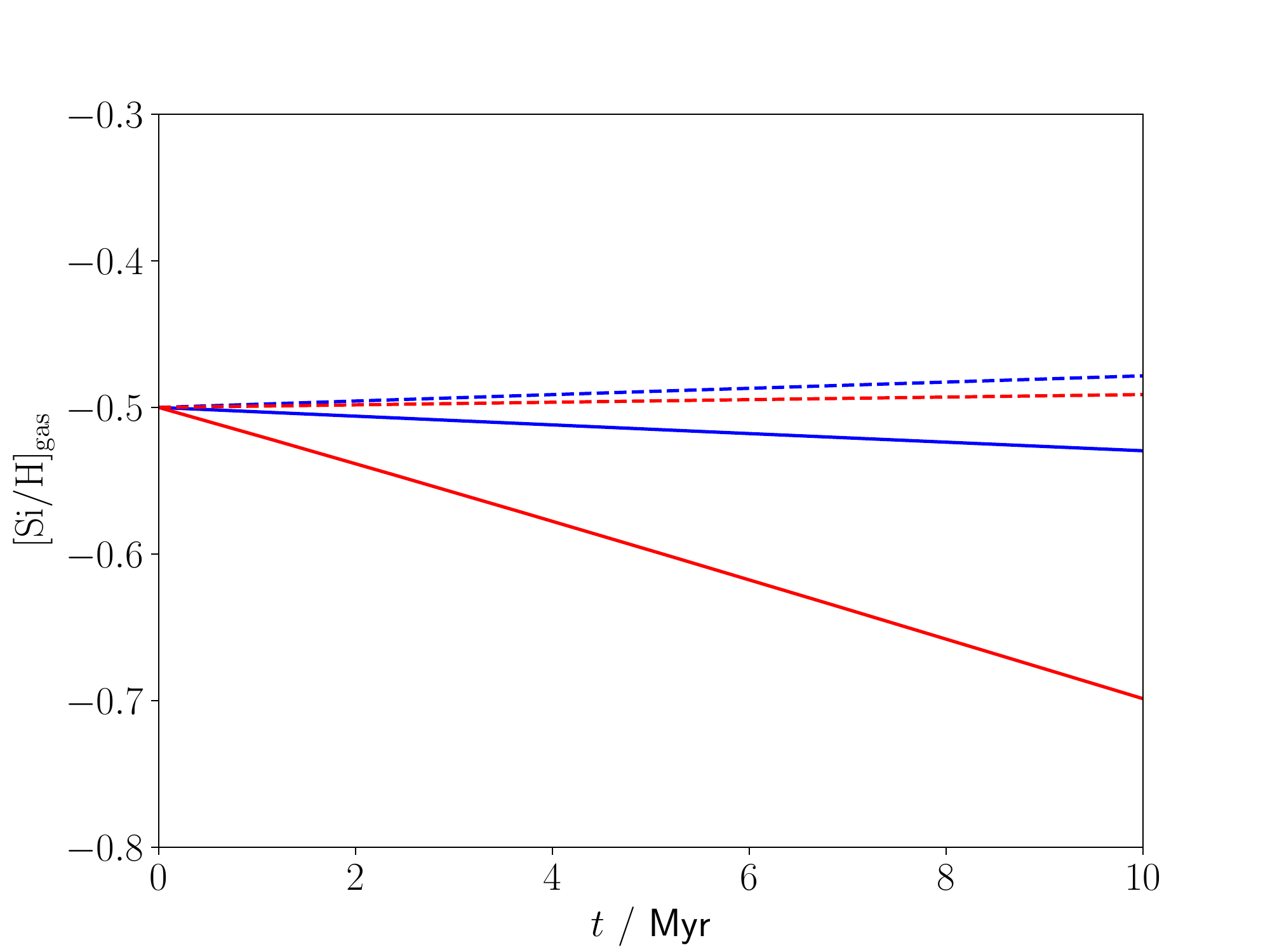}}
  \caption{\textit{Left:} Coulomb focusing factor for $\nh = 30 \pcc$, $T = 100 \kel$ and $x_e = 0.0015$ (solid lines) and $1.5 \times 10^{-4}$ (dashed lines). \textit{Right:} Silicon depletion versus time for $\nh = 30 \pcc$, $T = 100 \kel$ and $x_e = 0.0015$ (solid lines) and $1.5 \times 10^{-4}$ (dashed lines), with (blue) and without (red) an evolving size distribution.}
  \label{fig:lowe}
\end{figure*}

Accounting for grain-assisted recombination, the electron fraction of the CNM is reduced from the $0.0015$ value used by \citet{weingartner1999} by almost an order of magnitude, down to approximately the elemental abundance of carbon (the most common singly-ionised species under these conditions). Figure \ref{fig:lowe} shows the effect of reducing $x_e$ by a factor of ten on our fiducial model ($\nh = 30 \pcc$, $T = 100 \kel$). The Coulomb focusing factor is reduced for all grain radii, and the transition radius where it falls below unity occurs for smaller grains. This reduces the efficiency of grain growth to the point where there is no period of increasing depletion, and for an initial $\depsi = -0.5$ material is returned to the gas phase at all times. The \citet{zhukovska2016} model is thus not only overestimating the grain growth efficiency by neglecting the evolution of the size distribution, but also by adopting a potentially inaccurate set of physical conditions for the CNM.


\bsp	
\label{lastpage}
\end{document}